\documentclass[epj]{svjour}

\usepackage{graphicx}
\usepackage{fancyhdr}

\def\mytitle{My title} 
\def\myauthors{My name}  
\def\mytype{My type of session}
\def\mysession{My session}

\def\mytitle{Inhomogeneous preheating} 
\def\myauthors{Tomohiro Matsuda}    
\def\mytype{Parallel}    
\def\mysession{Cosmology and Astrophysics}

\begin{document}
\title{Inhomogeneous preheating in multi-field
models of cosmological perturbation}
\author{Tomohiro Matsuda\inst{1}
\thanks{\emph{Email:} matsuda@sit.ac.jp}%
\thanks{\emph{Present address:} Fusaiji, Okabe-machi, Saitama, Japan}%
}                     
%
%
\institute{Saitama Institute of Technology
}
%
\date{}
\abstract{We consider inhomogeneous preheating in multi-field models of
cosmological perturbation. After preheating, two fields are trapped at
an enhanced symmetric point. One field is an oscillating field and the
other is a light field that plays an important role in generating
perturbation. In this presentation, we consider two types of potential
for the light field. Unlike the usual modulated (p)reheating scenario,
there is no moduli problem because moduli-dependent couplings are not
needed. Since there is no moduli problem the inflation scale can
be lowered. 
} 
\maketitle
%

\section{Introduction}
\label{intro}
\begin{figure}[ht]
 \begin{center}
 \begin{picture}(450,110)(0,0)
 \resizebox{6cm}{!}{\includegraphics{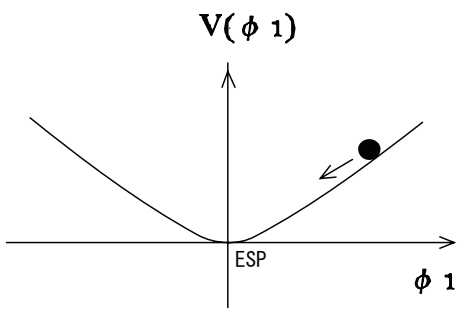}} 
 \end{picture}
\label{fig:fig424}
 \end{center}
\caption{Oscillation after inflation that induces preheating at the
 ESP.}
\label{pre-pic}
\end{figure}

\begin{figure}[ht]
 \begin{picture}(750,60)(0,0)
 \resizebox{8cm}{!}{\includegraphics{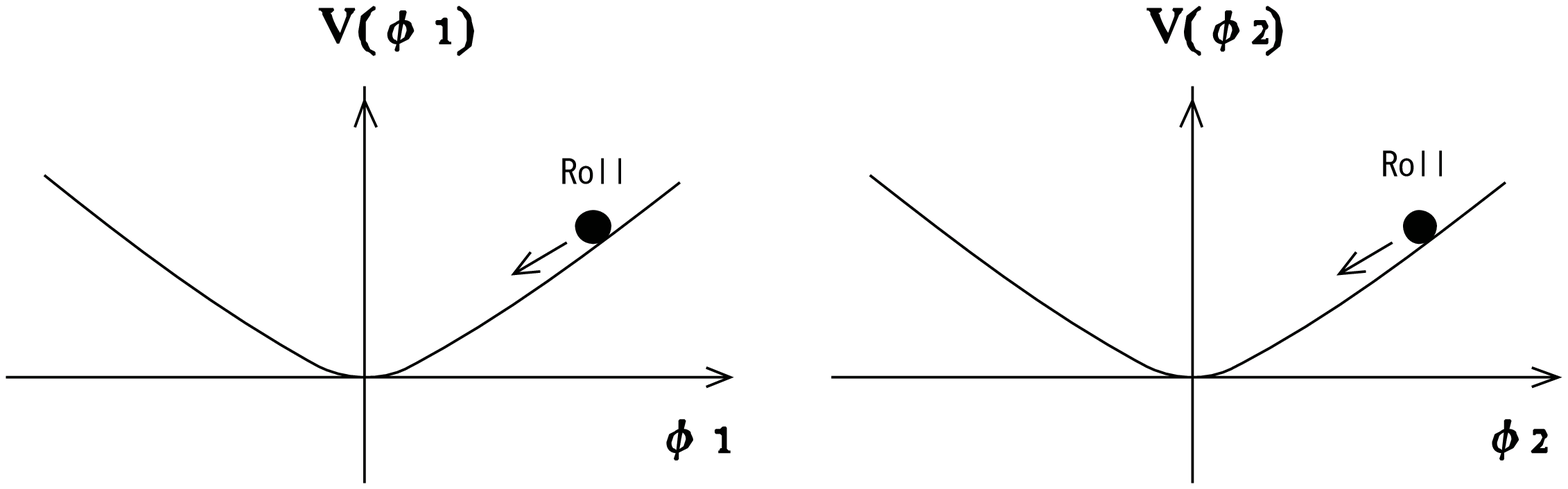}} 
 \end{picture}
\caption{Symmetric potential.}
\label{sym}
\end{figure}

\begin{figure}[ht]
 \begin{picture}(750,60)(0,0)
 \resizebox{8cm}{!}{\includegraphics{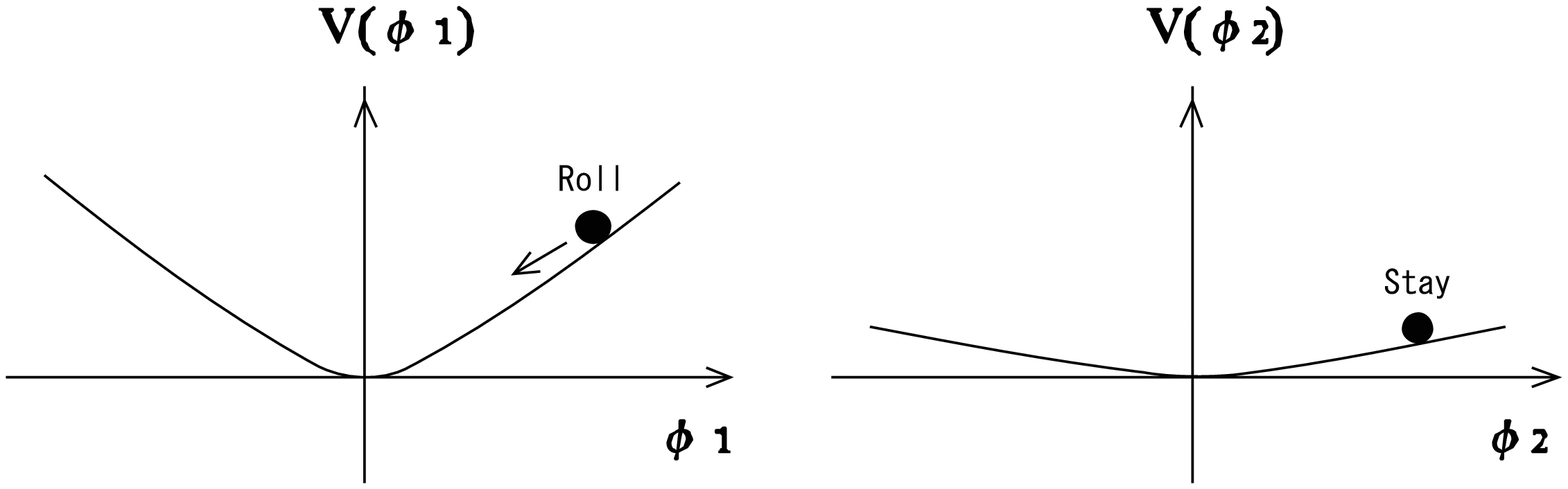}} 
 \end{picture}
\caption{Potential with a hierarchical mass difference.}
\label{asym}
\end{figure}

{\bf Motivation for the ``Alternatives'' to the traditional inflation}\\
We consider the following:
\begin{itemize}
\item \underline{Traditional Inflaton}\\
According to the traditional inflationary scenario, the spectrum of the
curvature perturbation is generated by the inflaton field. 
The spectrum is determined by the inflation model
      alone.  
\item \underline{Alternatives to the traditional inflation}\\
The primordial density perturbation may instead originate from the
      vacuum fluctuation of a ``non-inflaton'' field.  
\end{itemize}
Considering alternatives to the traditional scenario, the inflation
model can be eliminated as the source of the generation of the curvature
perturbation. 
Hence, we can expect that inflation may be separated from
the problems related to the generation of the cosmological
perturbation. 
For the treatment of nontraditional cosmological perturbation, we
introduce ``$\phi_2$'' as the non-inflaton field.\\
{\bf Problem in low-scale inflation (typical)}\\
For traditional inflation, the spectrum of the curvature perturbation is
given by 
\begin{equation}
{\cal P}_{\cal R}(k)=\frac{1}{24\pi^2 M_p^4}\frac{V_I}{\epsilon_I}
\quad{\textrm{(Traditional \,\,Inflation)}}
\end{equation}
Assuming that the scale of the inflation is much smaller than the Planck
scale, serious fine-tuning is required. 
Since a low-scale gravity model is still an important possibility that
could be tested in future experiments, it is important to consider how
low-scale inflation takes place. 
Future experiments (like BH production in LHC) may or may not put an
indirect bound on $H_I^{1/4}$, such as $H_I \le {\cal O}$(TeV).\\
{\bf Examples of the ``Alternatives''}
\begin{itemize}
\item {Curvatons\cite{curvaton1,matsuda_curvaton,topological_curv}} 
\item {Modulated (p)reheating\cite{Kofman,Inho_Reh_Dvali,alt2}} 
\item {Inhomogeneous preheating\cite{SSB-inst,matsuda_inst}}
\item {Generating $\delta N_e$ at the end of
      inflation\cite{alternate,fast-string}}  
\item Others(equally important)
\end{itemize}
In this talk, we mainly consider inhomogeneous preheating
 combined with curvatons or $\delta N_e$ generation at the end of
 inflation. 
We do not consider either moduli-dependent couplings nor fluctuation of
moduli that may lead to serious moduli problem.\\
{\bf What is inhomogeneous preheating?}\\
To explain inhomogeneous preheating, we consider first 
simple preheating and then introduce $\phi_2$ field to develop 
``multi-field preheating''.
We then explain how multi-field preheating induces inhomogeneous
preheating. 
Finally, we discuss generation of the cosmological perturbation
from inhomogeneous preheating combined with curvatons or
$\delta N_e$ generation at the end of inflation. 
Preheating\cite{original-PR} is induced by an oscillating field
$\phi_1$ (usually an inflaton).(See Fig.\ref{pre-pic})
We assume an interaction given by
\begin{equation}
{\cal L}= -\frac{1}{2}g^2 |\phi_1|^2 \chi^2,
\end{equation}
which induces a mass term for the preheat field $\chi$.
At the enhanced symmetric point (ESP), the effective mass of the preheat
field vanishes and non-adiabatic excitation of $\chi$ occurs, 
which induces efficient generation of the preheat field $\chi$.
The number density of the preheat field ($n_\chi$) that is
generated at the first scattering is given by
\begin{equation}
n_\chi = \frac{(g|\dot{\phi_1}(t_*)|)^{3/2}}{8\pi^3}
\exp\left[-\frac{\pi m_\chi^2}{g |\dot{\phi_1}(t_*)|}
\right].
\end{equation}
{\bf Multi-field extension}\\
In addition to the oscillating field $\phi_1$, we may add $\phi_2$ that
has the same coupling as $\phi_1$. 
This leads to multi-field preheating.
If the potential $V(\phi_1,\phi_2)$ is symmetric for rotation in
$(\phi_1, \phi_2)$ space, the potential looks like that shown in Fig.\ref{sym}.
On the other hand, if the global symmetry is badly broken (i.e., there
is a hierarchical mass difference $m_1 \gg m_2$), the potential looks
like that in Fig.\ref{asym}.
Depending on the situation, the trajectory of the multi-field
oscillation becomes 
\begin{itemize}
\item ``A straight line that precisely hits the ESP'' for a symmetric
      potential. (first picture in Fig.4)
\item ``A curved line that does not hit the ESP'' for slightly broken
      symmetry. (lower left in Fig.4)
\item ``An almost straight line that does not hit the ESP'' for a
      hierarchical mass difference. (right-hand side in Fig.4)
\end{itemize}
\begin{figure}[ht]
 \begin{center}
 \begin{picture}(750,140)(0,0)
 \resizebox{8cm}{!}{\includegraphics{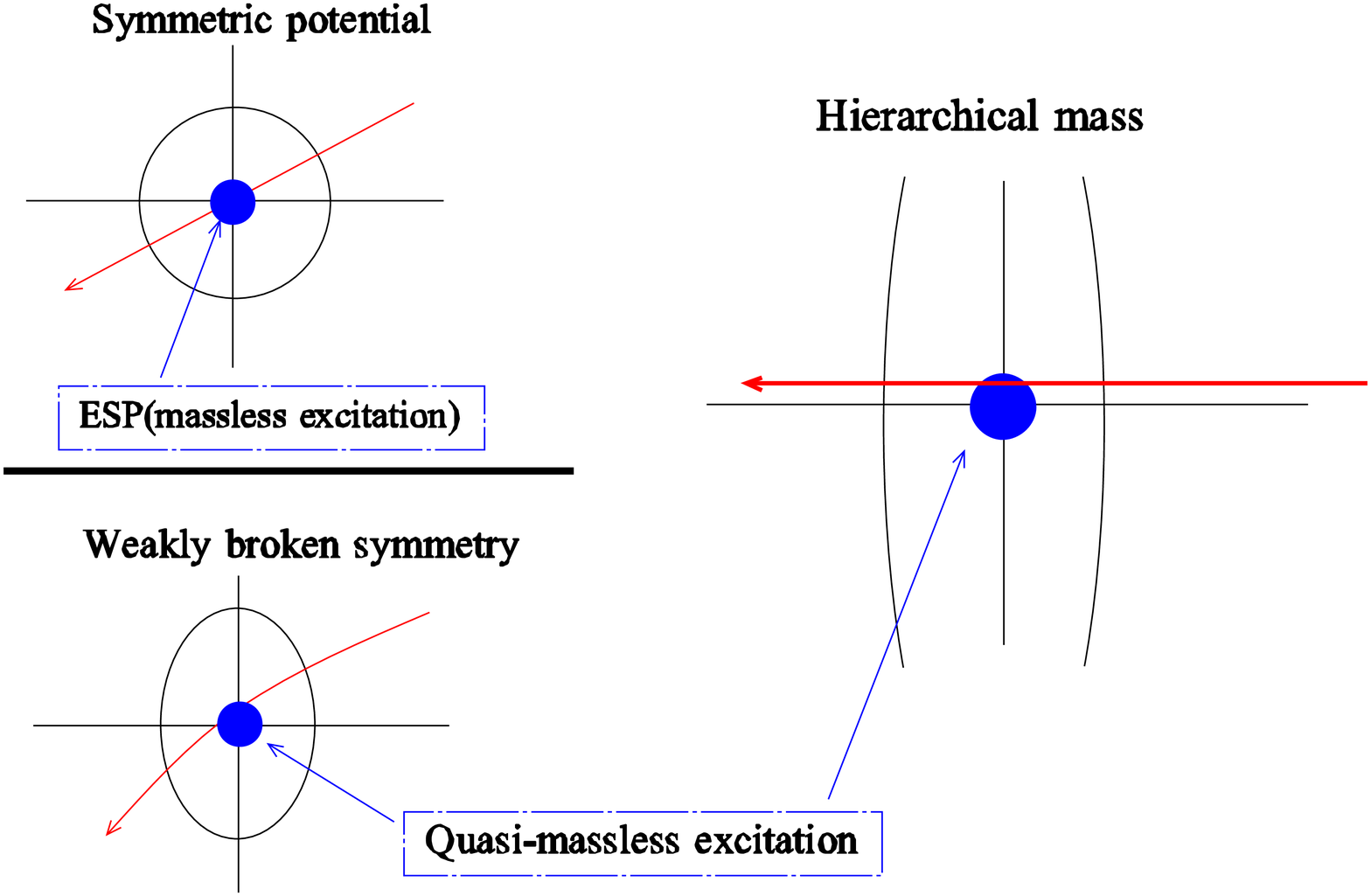}} 
 \end{picture}
 \end{center}
\caption{Trajectories of inflaton motion.}
\label{traj}
\end{figure}

\section{Inhomogeneous preheating; origin of the fluctuation in
  multi-field preheating}
\label{sec1}
Except for the symmetric potential, for which the trajectory precisely
hits the origin, the effective mass ($m_\chi$) does not vanish during
preheating. 
For a potential with a slightly broken symmetry, $m_\chi$ depends on
the initial angle and is given by $m_\chi(\theta)$.
For a potential with hierarchical masses, $m_\chi$ depends on
the initial value of the light field $\phi_2$ and is given by
$m_\chi(\phi_2)$. 
Therefore, 
\begin{itemize}
\item For the ``slightly broken symmetry'' the trajectory is determined
      by the initial angular parameter $\theta$. The origin of the
      fluctuation is denoted by $\delta \theta$, where $\theta$ is the
      $U(1)$ angle of the symmetry. 
\item For ``hierarchical mass'' the distance from the ESP is
      determined by the initial value of $\phi_2$. The origin of
      the fluctuation is denoted by $\delta \phi_2$, where $\phi_2$ is
      the additional light field of multi-field preheating.
\end{itemize}
In the hierarchical mass model, the light field $\phi_2$ gives mass
to the preheat field at the ESP:
\begin{equation}
m_{\chi}|_{ESP} \simeq g \phi_2.
\end{equation}
Therefore, the primordial fluctuation $\delta \phi_2$ leads
 to fluctuation of the mass $\delta m_\chi$, which finally
 induces inhomogeneous preheating and $\delta n_\chi\ne0$.
It is important to note that both the magnitude and typical length scale
of the fluctuation $\delta \phi_2$ is determined by the primordial
inflation. \\
{\bf Previous approaches}\\
In the following earlier previous approaches, ``instant decay'' has been
assumed for the preheat field: 
\begin{itemize}
\item Slightly broken symmetry\cite{SSB-inst}\\
{\it E. W. Kolb, A. Riotto, A. Vallinotto}.\\
{\bf The origin of the fluctuation is {$\delta \theta$}.}
\item Hierarchical mass difference(badly broken
      symmetry)\cite{matsuda_inst}
{\it T. Matsuda}.\\
{\bf The origin of the fluctuation is $\delta \phi_2$.}
\end{itemize}
For instant decay, generation of the cosmological perturbation and
reheating occurs just after the inhomogeneous preheating. This requires 
\begin{equation}
\left.\frac{\rho_\chi}{\rho_{total}}\right|_{ini}\sim 1,
\end{equation}
which puts a lower bound on the coupling constant.

\begin{figure}[ht]
 \begin{picture}(200,75)(0,0)
 \resizebox{8.5cm}{!}{\includegraphics{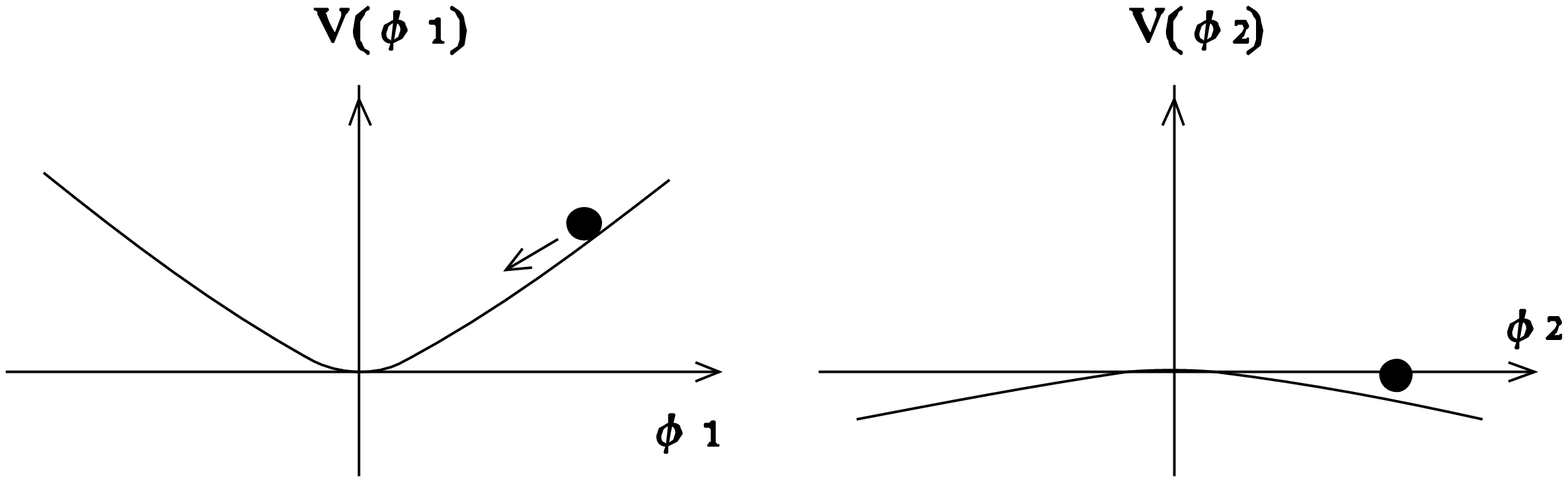}} 
 \end{picture}
\caption{Potential for trapping inflation after inhomogeneous preheating}
\label{FIGweak}
\end{figure}

Now we consider what happens if $\chi$ does not decay
instantaneously and whether it is possible to remove the condition
$\left.\frac{\rho_\chi}{\rho_{total}}\right|_{ini}\sim 1$ that leads to
$g \sim 1$ in the previous approaches.\\
{\bf Back reaction from the preheat field}\\
The effective potential induced by a stable $\chi$-field induces an 
attractive confining force to both $\phi_1$ and $\phi_2$. 
For single-field preheating, a similar situation has been discussed
for moduli trapping in string theory\cite{beauty_is}.
According to this treatment, the preheat field induces effective
potential 
\begin{equation}
V_c(\phi_i)\sim g  n_\chi |\phi_i|,
\end{equation}
which leads to an attractive confining force proportional to the
distance from the ESP.
This trapping can generate the cosmological perturbation from
inhomogeneous preheating. The obvious differences from the previous
approaches are (1) No instant decay is assumed (2)
The preheat field does not dominate the energy density just after
preheating.\\ 
{\bf \underline{Model 1} : Generating $\delta N_e$ at the end of trapping
inflation}\\
First, we discuss how to generate $\delta N_e$ at the end of 
trapping inflation\cite{beauty_is} with the potential
\begin{equation}
V(\phi_2) = -\frac{1}{2}m^2 \phi_2^2 +
 \lambda\frac{\phi_2^{n}}{M^{n-4}}.
\end{equation}
Note that the $\phi_2$-potential is inverted.(See Fig.\ref{FIGweak})

\begin{figure}[ht]
 \begin{picture}(200,100)(0,0)
 \resizebox{8.5cm}{!}{\includegraphics{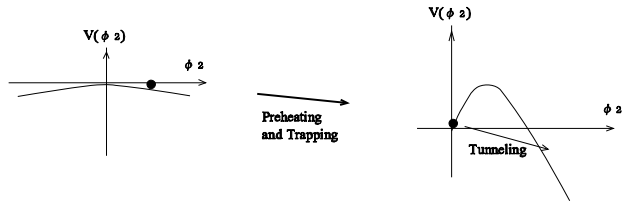}} 
 \end{picture}
\caption{Back reaction from the preheat field.}
\label{FIGback}
\end{figure}

This potential may remind the reader of thermal inflation induced by
thermal trapping. Note that in our model, trapping inflation is induced
by trapping after preheating but before reheating.  
Because of the back reaction from the preheat field, the effective
potential near the origin is significantly altered.(See
Fig.\ref{FIGback}.) 

\begin{figure}[ht]
 \begin{picture}(200,120)(0,0)
 \resizebox{9cm}{!}{\includegraphics{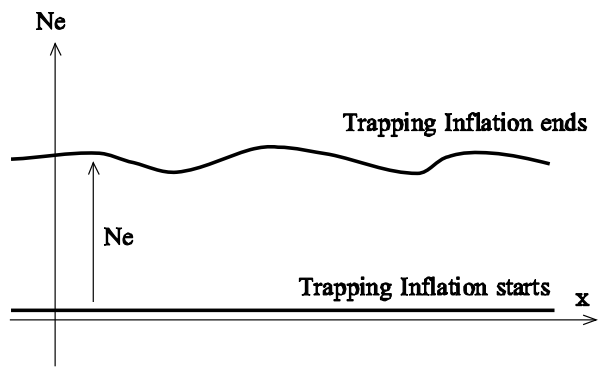}} 
 \end{picture}
\caption{Generating $\delta N_e$ from $\delta n_\chi$}
\label{gen}
\end{figure}

Let us summarize the process for generating the curvature perturbation
in this scenario. 
\begin{enumerate}
\item Preheating occurs due to $\phi_1$-oscillation
      while trapping occurs for both fields. 
\item $\phi_2$ is trapped at the local minimum at the origin.
\item The potential barrier $\Delta V$ decreases as 
      $\Delta V \propto  n_\chi^2$. 
\item Trapping inflation ends with the  $\phi_2$-tunneling.  
\end{enumerate}
{\bf Generating $\delta N_e$ from $\delta n_\chi$.}\\
Fig.\ref{gen} shows that the start-line of trapping inflation is independent
of the fluctuation $\delta n_\chi$ and is given by a flat 
  surface (the straight line at $N_e=0$).
On the other hand, the end-line is determined by the number density of
the preheat field $\chi$, which has the fluctuation $\delta n_\chi$. 
Note that trapping inflation is not the primary inflation but an
additional inflationary stage that starts after preheating. 

{\bf Calculation}\\
During trapping inflation, $V^{eff}(\phi_2)$ is given by 
\begin{equation}
V_2^{eff}(\phi_2) = V_0 -\frac{1}{2}m^2 \phi_2^2 +
\frac{\lambda|\phi_2|^{n_2}}{M_2^{n_2-4}} + g n_\chi |\phi_2|.
\end{equation}
For the effective potential near the origin, the effective potential for
$\phi_2>0$ is written as
\begin{equation}
V_2^{eff}(\phi_2) \simeq V_0 -\frac{1}{2}m^2 
\left(\phi_2 - \frac{g n_\chi}{m^2}\right)^2 +\frac{g^2 n_\chi^2}{2m^2}.
\end{equation}
The number of e-foldings that elapse during the trapping inflation is
given by 
\begin{equation}
N_e \sim \frac{1}{3}\ln\left(\frac{n_\chi(t_i)}{n_\chi(t_e)}\right).
\end{equation}
$\delta N_e$ generated at the end of inflation is
\begin{equation}
\delta N_e \sim \frac{g \phi_2 \delta \phi_2}{v},
\end{equation}
where $v$ is the velocity of the oscillating field at the ESP.
Our result shows that: (1) Low-scale inflation ($H_I\sim GeV$) is
successful. (2) The non-Gaussian parameter is always large, $|f_{NL}| >
1$.
Unfortunately, these results depend crucially on the initial condition.\\
{\bf \underline{Model 2} : Weak trapping and non-oscillating (NO) curvatons}\\
We consider the quintessential potential for $\phi_2$:
\begin{equation}
V(\phi_2) = 
 \frac{\Lambda}{(\phi_2)^n}.
\end{equation}
Note that the attractive force from the preheat field acts on $\phi_2$
 to prevent $\phi_2$ from rolling down the potential.
\begin{figure}[ht]
  \begin{picture}(350,40)(0,0)
 \resizebox{8.5cm}{!}{\includegraphics{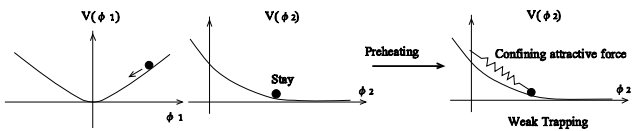}} 
  \end{picture}
\caption{Attractive force acting on $\phi_2$}
\end{figure}

Here we consider the situation
\begin{itemize}
\item  The preheat Field ($\chi$) is identified with the 
   curvaton.
\item  There is the back reaction from the preheat field (i.e., from the 
 curvaton).
\end{itemize}

The late-time evolution is found from the force-balance equation
\begin{equation}
g n_\chi(t) -\frac{n M^{n+4}}{\phi_2^{n+1}}=0,
\end{equation}
which leads to the evolution of the expectation value $\phi_2(t)$,
\begin{equation}
\label{rela_si}
\phi_2(t) =M\left(\frac{n M^{3}}{g n_\chi(t)}\right)^{1/(n+1)}.
\end{equation}
From these equations we can calculate the ratio of $\rho_\chi$
to $V(\phi_2)$,
\begin{equation}
\frac{\rho_{\chi}}{V(\phi_2)}=n.
\end{equation}
Since the number density $n_\chi$ evolves as $n_\chi\propto
a^{-3}$, the energy density of the preheat field $\rho_{\chi}$
evolves as
\begin{equation}
\rho_{\chi}\propto a^{-3(1-\frac{1}{n+1})}.
\end{equation}
Note that the mass of the curvaton $m_\chi$ grows as
\begin{equation}
m_\chi \simeq g \phi_2(t) \propto a^{\frac{3}{n+1}}.
\end{equation}
As a consequence, we find clear differences from the normal curvaton;
\begin{enumerate}  
\item  The time when the curvaton starts to oscillate is determined by the
       mass of the oscillating field $\phi_1$, which is 
       independent of the curvaton mass ($m_\chi$).  
       Note that $\phi_1$ may or may not be the inflaton.
\item  The time when the curvaton decays is determined by
       $m_\chi(t)$, which increases with time.
\item The density of the NO curvaton decreases slower than the
       matter density.
\end{enumerate}
As a result, the cosmological bound for the NO curvaton is very
different from the usual curvatons. 

Let us consider an example.
For the Quintessential potential
\begin{equation}
 V(\phi_2)=\frac{M^8}{(\phi_2)^4}, \quad \quad M=10^2 GeV 
\end{equation}
we obtain $T_{R} \simeq 1 $MeV.
There is no obvious bound for the Hubble parameter above $H_I\sim {\cal
O}$(GeV), but again, the results depend crucially on the initial
condition.\\
{\bf \underline{Conclusions and summary}}\\
We have presented two multi-field models of cosmological
perturbation. 
The first model deals with a typical double-well potential, which has
the same form as the one that has been used for thermal inflation. 
We showed that a combination of inhomogeneous preheating and
trapping inflation leads to the generation of the curvaton
perturbation. 
We conclude that inhomogeneous preheating is an interesting possibility
and that the traditional scenario for generating cosmological
perturbations can be replaced by our proposed alternatives. Future
cosmological observations should help to determine which of these
alternatives is the most suitable; non-Gaussianity in particular may be
a key observation.\footnote{See also Ref.\cite{hybrid}.}
However, a more efficient method is required to properly identify which
model is most appropriate, and we are currently conducting research in
this direction.
%
%

\end{document}